\documentclass[11pt]{article}

\usepackage{troffms}

\begin{document}

\footer{\rm\thepage}

\title{Solution of linearized Fokker - Planck equation \\
for incompressible fluid \\
}
\author{Igor A. Tanski \\\
       Moscow, Russia \\
       tanski.igor.arxiv@gmail.com \\
}
        
\rule{6in}{1pt}
        
\begin{abstract}
In this work we construct algebraic equation for elements of spectrum of linearized Fokker - Planck differential operator for incompressible fluid. We calculate roots of this equation using simple numeric method. For all these roots real part is positive, that is corresponding solutions are damping. Eigenfunctions of linearized Fokker - Planck differential operator for incompressible fluid are expressed as linear combinations of eigenfunctions of usual Fokker - Planck differential operator. Poisson's equation for pressure is derived from incompressibility condition. It is stated, that the pressure could be totally eliminated from dynamics equations. The Cauchy problem setup and solution method is presented. The role of zero pressure solutions as eigenfunctions for confluent eigenvalues is emphasized.
\end{abstract}

\shead{Keywords}

\par\noindent
Fokker-Planck equation, incompressible fluid, linear operator spectrum

\rule{6in}{1pt}

\section{Introduction}

\par
In our previous work [1] we derived linearized Fokker - Planck equation for incompressible fluid
$$
\int_V n dv_x dv_y dv_z = 0 .
\eqno (1)$$
$$
{\partial n   \over \partial t} +
v_j {\partial n   \over \partial x_j} - 
\alpha\  {\partial \over \partial v_j} (v_j n ) +
\left(
{\alpha \over k }\right)\ 
\left( {\alpha \over 2 \pi k} \right)^{3/2} 
\exp \left[
- {\alpha \over 2k }v_j v_j
\right]\ 
v_k
{\partial p   \over \partial x_k} 
= k\  {\partial^2 n   \over \partial v_j \partial v_j} .
\eqno (2)$$
\par\noindent
where
\par\noindent
$n = n(t, x_1 , x_2 , x_3 , v_1 , v_2 , v_3 )$ - density;
\par\noindent
$p = p(t, x_1 , x_2 , x_3 )$ - pressure;
\par\noindent
$t$ - time variable;
\par\noindent
$x_1 , x_2 , x_3 $ - space coordinates;
\par\noindent
$v_1 , v_2 , v_3$ - velocities;
\par\noindent
$\alpha$ - coefficient of damping;
\par\noindent
$k$ - coefficient of diffusion.
\par
No attempt was made to solve this equation.
\par
Peculiarity of this equation consists in the fact, that for two unknown variables $n$ and $p$ we have only one differential equation. This is enough, because there is additional normalization requirement (1) on $n$ variable and $p$ variable depends only on space coordinates $x, y, z$ and time $t$.
\par
In [2] some simple solutions of nonlinear equation were studied. Particularly flow with zero pressure is of interest for our present studies, because for this flow the source of nonlinearity is absent. This flow solves both nonlinear and linear equations.
\par
Present work is devoted to solution of linearized equation. We consider only the case of parallelepiped with opposite sides identified (i.e. "periodic boundary conditions"). We try to formulate and solve Cauchy problem for linearized equation.
\section{Fourier decomposition of solution}
\par
We know the form of Cauchy problem solution for the case of usual Fokker-Planck equation (see [3]).
$$
n (t, x_j , v_j ) =
\sum_{ m_1 = - \infty }^{ + \infty }
\sum_{ m_2 = - \infty }^{ + \infty }
\sum_{ m_3 = - \infty }^{ + \infty }
\sum_{ p_1 = 0 }^{ \infty }
\sum_{ p_2 = 0 }^{ \infty }
\sum_{ p_3 = 0 }^{ \infty }
A_{ m_1 m_2 m_3 p_1 p_2 p_3 } (t)\  \phi_{ m_1 m_2 m_3 p_1 p_2 p_3 } ,
\eqno (3)$$

\par\noindent
where eigenfunctions of usual Fokker - Planck operator are

$$
\phi_{ m_1 m_2 m_3 n_1 n_2 n_3 } =
\prod_{{j=1}}^{{j=3}}
\exp 
\left( 
2 \pi i {m_j \over a_j }( x_j - {v_j \over \alpha }) 
\right)
\exp
\left(
- {\alpha \over 2k } v_j^2
\right) 
H_{{n}_j}
\left(
\sqrt {\alpha \over { 2k }}
\left(
v_j +
{ {4 \pi i m_j k } \over { \alpha^2 a_j } }
\right)
\right)
 .
\eqno (4)$$
\par
It is only natural to seek solution of the present problem in the same form. We need only add expression for the new $p$ variable
$$
p(t, x_j ) =
\sum_{ m_1 = - \infty }^{ + \infty }
\sum_{ m_2 = - \infty }^{ + \infty }
\sum_{ m_3 = - \infty }^{ + \infty }
P_{ m_1 m_2 m_3 } (t)\  
\prod_{{j=1}}^{{j=3}}
\exp 
\left( 
2 \pi i {m_j \over a_j }x_j 
\right) ,
\eqno (5)$$
\par
We try to represent the coefficient before ${\partial p   \over \partial x_k} $ in equation (2) in the same way as a sum of Fourier series
$$
\left(
{\alpha \over k }\right)\ 
\left( {\alpha \over 2 \pi k} \right)^{3/2} 
\prod_{{j=1}}^{{j=3}}
\exp 
\left( 
2 \pi i {m_j \over a_j }x_j 
\right)
\exp \left[
- {\alpha \over 2k }v_j v_j
\right]\ 
v_k =
\eqno (6)$$
$$
=
\sum_{ p_1 = 0 }^{ \infty }
\sum_{ p_2 = 0 }^{ \infty }
\sum_{ p_3 = 0 }^{ \infty }
B_{ m_1 m_2 m_3 n_1 n_2 n_3 } (k) \phi_{ m_1 m_2 m_3 n_1 n_2 n_3 } .
$$
\par
In expressions (3, 5, 6) we introduced following Fourier coefficients:
\par\noindent
$A_{ m_1 m_2 m_3 p_1 p_2 p_3 } (t)$ - coefficients of decomposition unknown variable $n$;
\par\noindent
$P_{ m_1 m_2 m_3 } (t)$ - coefficients of decomposition of unknown variable $p$;
\par\noindent
$B_{ m_1 m_2 m_3 n_1 n_2 n_3 } (k)$ - coefficients of decomposition of known variable, which represent coefficient before $p$ gradient. Values of $B$ are presented below.
\par
Using these coefficients, we rewrite equation (2) as
$$
{d \over dt }A_{ m_1 m_2 m_3 n_1 n_2 n_3 } (t)
+ 
\sum_{{j=1}}^{{j=3}}
\left[
\alpha n_j +
k\  \left(
{2 \pi m_j  \over \alpha a_j }
\right)^2
\right]\ 
A_{ m_1 m_2 m_3 n_1 n_2 n_3 } (t)
=
\eqno (7)$$
$$
=
2 \pi i {m_k \over a_k }B_{ m_1 m_2 m_3 n_1 n_2 n_3 } (k)
P_{ m_1 m_2 m_3 } (t)\  
 .
$$
\par
We reduced the partial differential equation (2) to system of ordinary differential equations for Fourier coefficients. To proceed with solution, we need expressions for known coefficients $B_{ m_1 m_2 m_3 n_1 n_2 n_3 } (k)$.
\section{Expressions for Fourier coefficients}
\par
In this auxiliary section we find explicit expressions for Fourier coefficients of some known functions of velocities. These functions are products of multipliers, which depend only on one independent variable. Therefore we can consider only one velocity variable in this section.
\par
We start from definitions
$$
\phi_{mn} = 
\exp \left(
- { {2 \pi i m}  \over {\alpha a} } v
\right)\ 
\exp
\left(
- {\alpha \over {2k} } v^2
\right)
H_n 
\left(
\sqrt {\alpha \over { 2k } }
\left(
v +
{ {4 \pi i m k}  \over {\alpha^2 a }}
\right)
\right)
 .
\eqno (8)$$
$$
\psi_{mn} = 
\exp \left(
- { {2 \pi i m}  \over {\alpha a} } v
\right)\ 
H_n 
\left(
\sqrt {\alpha \over { 2k }}
\left(
v +
{{ 4 \pi i m k} \over {\alpha^2 a} }
\right)
\right)
 .
\eqno (9)$$
\par
Functions $\phi_{mn}$ and $\psi_{mn}$ are of course orthogonal (see [3])
$$
\int_{{-} \infty}^{\infty}
\phi_{mp}\ \  \psi_{mq} dv =
\exp
\left[
- {\alpha \over 2k }\ 
\left( 
{4 \pi m k  \over \alpha^2 a} 
\right)^2
\right]\ 
\sqrt {{2 \pi k  \over \alpha }} 
\delta_{pq} (-2)^p p! .
\eqno (10)$$
\par
Let us find Fourier coefficients for following functions:
$$
\exp \left[
- {\alpha \over 2k }v^2
\right] =
\sum_{n=0}^{\infty}
a_{mn} \phi_{mn} .
\eqno (11)$$
$$
\exp \left[
- {\alpha \over 2k }v^2
\right]\ 
v =
\sum_{n=0}^{\infty}
b_{mn} \phi_{mn} .
\eqno (12)$$
\par
To find these coefficients, we need to calculate integrals
$$
a_{mn} = 
\exp
\left[
{\alpha \over 2k }\ 
\left( 
{4 \pi m k  \over \alpha^2 a} 
\right)^2
\right]\ 
\sqrt {{\alpha \over 2 \pi k}} 
{1 \over  (-2)^n n! }
\int_{{-} \infty}^{\infty}
\exp \left[
- {\alpha \over 2k }v^2
\right]\ \  \psi_{mn} dv .
\eqno (13)$$
$$
b_{mn} = 
\exp
\left[
{\alpha \over 2k }\ 
\left( 
{4 \pi m k  \over \alpha^2 a} 
\right)^2
\right]\ 
\sqrt {{\alpha \over 2 \pi k}} 
{1 \over  (-2)^n n! }
\int_{{-} \infty}^{\infty}
\exp \left[
- {\alpha \over 2k }v^2
\right]\  v\  \psi_{mn} dv .
\eqno (14)$$
\par
We calculated these integrals in our previous work [2]
$$
\int_{{-} \infty}^{\infty}
\exp \left[
- {\alpha \over 2k }v^2
\right]\  \psi_{mn} dv =
\eqno (15)$$
$$
=
\int_{{-} \infty}^{\infty}
\exp \left[
- {\alpha \over 2k }v^2
\right]\  \exp \left(
- {2 \pi i m  \over \alpha a} v
\right)\ 
H_n 
\left(
\sqrt {{\alpha \over 2k }}
\left(
v +
{4 \pi i m k  \over \alpha^2 a}
\right)
\right)\  dv =
$$
$$
=
\sqrt {{2 \pi k  \over a }}
\left(
{2k \over \alpha }\right)^{n/2}
\exp \left[
- {k \over 2 \alpha}
\left(
{2 \pi m  \over \alpha a}
\right)^2
\right]\ 
\left(
{2 \pi i m  \over \alpha a}
\right)^n .
$$
$$
\int_{{-} \infty}^{\infty}
\exp \left[
- {\alpha \over 2k }v^2
\right]\  v\  \psi_{mn} dv =
\eqno (16)$$
$$
=
\int_{{-} \infty}^{\infty}
\exp \left[
- {\alpha \over 2k }v^2
\right]\  v\  \exp \left(
- {2 \pi i m  \over \alpha a} v
\right)\ 
H_n 
\left(
\sqrt {{\alpha \over 2k }}
\left(
v +
{4 \pi i m k  \over \alpha^2 a}
\right)
\right)\  dv =
$$
$$
=
\sqrt {{2 \pi k  \over a }}
\left(
{2k \over \alpha }\right)^{n/2}
\exp \left[
- {k \over 2 \alpha}
\left(
{2 \pi m  \over \alpha a}
\right)^2
\right]\ 
\left[
-
{k \over \alpha} 
\left( 
{ 2 \pi i m   \over \alpha a}
\right)^{n+1} + 
n\ 
\left( 
{ 2 \pi i m   \over \alpha a}
\right)^{n-1}
\right]
.
$$
\par
Thus we get following expressions for Fourier coefficients

\boxit{
$$
a_{mn} = 
\exp \left[
6 {k \over \alpha }\left(
{\pi m  \over \alpha a}
\right)^2
\right]\ 
{1 \over  2^n n! }
\left(
{2k \over \alpha }\right)^{n/2}
\left(
{2 \pi i m  \over \alpha a}
\right)^n
 .
\eqno (17)$$
}

\boxit{
$$
b_{mn} = 
\exp \left[
6 {k \over \alpha }\left(
{\pi m  \over \alpha a}
\right)^2
\right]\ 
{1 \over  2^n n! }
\left(
{2k \over \alpha }\right)^{n/2}
\left[
-
{k \over \alpha} 
\left( 
{ 2 \pi i m   \over \alpha a}
\right)^{n+1} + 
n\ 
\left( 
{ 2 \pi i m   \over \alpha a}
\right)^{n-1}
\right] .
\eqno (18)$$
}
\section{Dynamics of Fourier coefficients}
\par
In this section we substitute expressions for known coefficients in terms of $a_{{m}_2 n_2}$ and $b_{{m}_1 n_1}$ coefficients, which we find in the last section, to the main equation (7). Namely, we use expressions
$$
B_{ m_1 m_2 m_3 n_1 n_2 n_3 } (1) =
\left(
{\alpha \over k }\right)\ 
\left( {\alpha \over 2 \pi k} \right)^{3/2} 
b_{{m}_1 n_1}
a_{{m}_2 n_2}
a_{{m}_3 n_3} .
\eqno (19)$$
$$
B_{ m_1 m_2 m_3 n_1 n_2 n_3 } (2) =
\left(
{\alpha \over k }\right)\ 
\left( {\alpha \over 2 \pi k} \right)^{3/2} 
a_{{m}_1 n_1}
b_{{m}_2 n_2}
a_{{m}_3 n_3} .
\eqno (20)$$
$$
B_{ m_1 m_2 m_3 n_1 n_2 n_3 } (3) =
\left(
{\alpha \over k }\right)\ 
\left( {\alpha \over 2 \pi k} \right)^{3/2} 
a_{{m}_1 n_1}
a_{{m}_2 n_2}
b_{{m}_3 n_3} .
\eqno (21)$$
\par
Then (7) reads
$$
{d \over dt }A_{ m_1 m_2 m_3 n_1 n_2 n_3 } (t)
+ 
\sum_{{j=1}}^{{j=3}}
\left[
\alpha n_j +
k\  \left(
{2 \pi m_j  \over \alpha a_j }
\right)^2
\right]\ 
A_{ m_1 m_2 m_3 n_1 n_2 n_3 } (t)
=
\eqno (22)$$
$$
=
2 \pi i\ 
\left(
{\alpha \over k }\right)\ 
\left( {\alpha \over 2 \pi k} \right)^{3/2} 
\left(
{m_1 \over a_1 }b_{{m}_1 n_1}
a_{{m}_2 n_2}
a_{{m}_3 n_3} +
{m_2 \over a_2 }a_{{m}_1 n_1}
b_{{m}_2 n_2}
a_{{m}_3 n_3} +
\right.
$$
$$
+
\left.
{m_3 \over a_3 }a_{{m}_1 n_1}
a_{{m}_2 n_2}
b_{{m}_3 n_3}
\right)\ 
P_{ m_1 m_2 m_3 } (t)\  
 .
$$
\par\noindent
or
$$
{d \over dt }A_{ m_1 m_2 m_3 n_1 n_2 n_3 } (t)
+ 
\sum_{{j=1}}^{{j=3}}
\left[
\alpha n_j +
k\  \left(
{2 \pi m_j  \over \alpha a_j }
\right)^2
\right]\ 
A_{ m_1 m_2 m_3 n_1 n_2 n_3 } (t)
=
\eqno (23)$$
$$
=
2 \pi i\ 
\left(
{\alpha \over k }\right)\ 
\left( {\alpha \over 2 \pi k} \right)^{3/2} 
a_{{m}_1 n_1}
a_{{m}_2 n_2}
a_{{m}_3 n_3}\ 
\left(
{m_1 \over a_1 }b_{{m}_1 n_1 
\over a_{ }m_1 n_1} +
{m_2 \over a_2 }b_{{m}_2 n_2 
\over a_{ }m_2 n_2}
\right.
 +
$$
$$
+
\left.
{m_3 \over a_3 }b_{{m}_3 n_3 
\over a_{ }m_3 n_3}
\right)\ 
P_{ m_1 m_2 m_3 } (t)\  
 .
$$
\par
This is main equation, which describe dynamics of Fourier coefficients. We delay actual substitution of $a_{{m}_i n_j}$ and $b_{{m}_i n_j}$ until (31).
\section{Incompressibility condition}
\par
We use the incompressibility condition (1) to eliminate $P_{ m_1 m_2 m_3 } (t) $ from (23).
\par
(1) and (3) imply
$$
\sum_{ n_1 = 0 }^{ \infty }
\sum_{ n_2 = 0 }^{ \infty }
\sum_{ n_3 = 0 }^{ \infty }
A_{ m_1 m_2 m_3 n_1 n_2 n_3 } (t)\  
\int_V
\phi_{ m_1 m_2 m_3 n_1 n_2 n_3 } 
dv_x dv_y dv_z
= 0 .
\eqno (24)$$
\par
Let us denote
$$
c_{mn} = 
\int_{{-} \infty}^{\infty}
\exp \left[
- {\alpha \over 2k }v^2
\right]\  \psi_{mn} dv =
\int_{{-} \infty}^{\infty}
\phi_{mn} dv =
\eqno (25)$$
$$
=
\sqrt {{2 \pi k  \over a }}
\left(
{2k \over \alpha }\right)^{n/2}
\exp \left[
- {k \over 2 \alpha}
\left(
{2 \pi m  \over \alpha a}
\right)^2
\right]\ 
\left(
{2 \pi i m  \over \alpha a}
\right)^n .
$$
\par
So incompressibility condition is equivalent to following equation
$$
\sum_{ n_1 = 0 }^{ \infty }
\sum_{ n_2 = 0 }^{ \infty }
\sum_{ n_3 = 0 }^{ \infty }
c_{{m}_1 n_1}
c_{{m}_2 n_2}
c_{{m}_3 n_3}
A_{ m_1 m_2 m_3 n_1 n_2 n_3 } (t)\  
= 0 .
\eqno (26)$$
\par
Let us suppose, that coefficients $A_{ m_1 m_2 m_3 n_1 n_2 n_3 } (t)$ satisfy (26) at the moment $t$. They must satisfy this equation at the next moment $t+dt$. Values of $A$ at the next moment are defined by dynamics equation (23), which contains besides $A(t)$ also pressure $P$. Therefore incompressibility condition, written for the next moment, will give us equation for pressure $P$. Derivation of this equation is rather long procedure, which ends in equation (42).
\par
Differentiate (26) on time $t$ and get
$$
\sum_{ n_1 = 0 }^{ \infty }
\sum_{ n_2 = 0 }^{ \infty }
\sum_{ n_3 = 0 }^{ \infty }
c_{{m}_1 n_1}
c_{{m}_2 n_2}
c_{{m}_3 n_3}
A'_{ m_1 m_2 m_3 n_1 n_2 n_3 } (t)\  
= 0 .
\eqno (27)$$
\par
Let us find $A'_{ m_1 m_2 m_3 n_1 n_2 n_3 } (t)$ from (22) and substitute this value to (27)
$$
\sum_{ n_1 = 0 }^{ \infty }
\sum_{ n_2 = 0 }^{ \infty }
\sum_{ n_3 = 0 }^{ \infty }
c_{{m}_1 n_1}
c_{{m}_2 n_2}
c_{{m}_3 n_3}
\sum_{{j=1}}^{{j=3}}
\left[
\alpha n_j +
k\  \left(
{2 \pi m_j  \over \alpha a_j }
\right)^2
\right]\ 
A_{ m_1 m_2 m_3 n_1 n_2 n_3 } (t)
=
\eqno (28)$$
$$
=
2 \pi i\ 
\left(
{\alpha \over k }\right)\ 
\left( {\alpha \over 2 \pi k} \right)^{3/2} 
\sum_{ n_1 = 0 }^{ \infty }
\sum_{ n_2 = 0 }^{ \infty }
\sum_{ n_3 = 0 }^{ \infty }
c_{{m}_1 n_1}
c_{{m}_2 n_2}
c_{{m}_3 n_3}
\times
$$
$$
\times\ 
\left(
{m_1 \over a_1 }b_{{m}_1 n_1}
a_{{m}_2 n_2}
a_{{m}_3 n_3} +
{m_2 \over a_2 }a_{{m}_1 n_1}
b_{{m}_2 n_2}
a_{{m}_3 n_3} 
\right.
+
$$
$$
+
\left.
{m_3 \over a_3 }a_{{m}_1 n_1}
a_{{m}_2 n_2}
b_{{m}_3 n_3}
\right)\ 
P_{ m_1 m_2 m_3 } (t)\  
 .
$$
\par
Let us use (26) to remove term with $\left(
{2 \pi m_j  \over \alpha a_j }
\right)^2$
$$
\sum_{ n_1 = 0 }^{ \infty }
\sum_{ n_2 = 0 }^{ \infty }
\sum_{ n_3 = 0 }^{ \infty }
c_{{m}_1 n_1}
c_{{m}_2 n_2}
c_{{m}_3 n_3}
( n_1 + n_2 + n_3 )
A_{ m_1 m_2 m_3 n_1 n_2 n_3 } (t)
=
\eqno (29)$$
$$
=
2 \pi i\ 
\left(
{1 \over k }\right)\ 
\left( {\alpha \over 2 \pi k} \right)^{3/2} 
\sum_{ n_1 = 0 }^{ \infty }
\sum_{ n_2 = 0 }^{ \infty }
\sum_{ n_3 = 0 }^{ \infty }
c_{{m}_1 n_1}
c_{{m}_2 n_2}
c_{{m}_3 n_3}
\times
$$
$$
\times\ 
\left(
{m_1 \over a_1 }b_{{m}_1 n_1}
a_{{m}_2 n_2}
a_{{m}_3 n_3} +
{m_2 \over a_2 }a_{{m}_1 n_1}
b_{{m}_2 n_2}
a_{{m}_3 n_3} 
\right.
+
$$
$$
+
\left.
{m_3 \over a_3 }a_{{m}_1 n_1}
a_{{m}_2 n_2}
b_{{m}_3 n_3}
\right)\ 
P_{ m_1 m_2 m_3 } (t)\  =
0
 .
$$
\par
We can easily calculate the sums on $n_i$ in RHS of (29). For this purpose let us introduce the partial sum on the group of terms with constant $(n_1 + n_2 + n_3 ) = J$.
$$
S_J
=
2 \pi i\ 
\left(
{\alpha \over k }\right)\ 
\left( {\alpha \over 2 \pi k} \right)^{3/2}
\sum_{ n_1 + n_2 + n_3 = J }
c_{{m}_1 n_1}
c_{{m}_2 n_2}
c_{{m}_3 n_3}
\times
\eqno (30)$$
$$
\times
a_{{m}_1 n_1}
a_{{m}_2 n_2}
a_{{m}_3 n_3}
\left(
{m_1 \over a_1 } {b_{m_1 n_1} \over a_{ m_1 n_1 }}
 +
{m_2 \over a_2 } {b_{m_2 n_2} \over a_{ m_2 n_2} }
 +
{m_3 \over a_3 } {b_{m_3 n_3} \over a_{ m_3 n_3 } }
\right)\  .
$$
\par
Let us substitute to (30) values of coefficients $a_{ij} , b_{ij} , c_{ij}$
$$
S_J
=
\left(
{\alpha^2  \over k }\right)\ 
\left( {\alpha \over 2 \pi k} \right)^{3/2}
\sqrt {{2 \pi k  \over a_1 }}
\sqrt {{2 \pi k  \over a_2 }}
\sqrt {{2 \pi k  \over a_3 }}
\left(
{k \over \alpha }\right)^J
\times
\eqno (31)$$
$$
\times
\exp \left[
{k \over \alpha }\left(
{2 \pi m_1  \over \alpha a_1}
\right)^2
\right]\ 
\exp \left[
{k \over \alpha }\left(
{2 \pi m_2  \over \alpha a_2}
\right)^2
\right]\ 
\exp \left[
{k \over \alpha }\left(
{2 \pi m_3  \over \alpha a_3}
\right)^2
\right]\ 
\times
$$
$$
\times
\left(
{k \over \alpha} 
\left[
\left( 
{ 2 \pi m_1   \over \alpha a_1}
\right)^2 + 
\left( 
{ 2 \pi m_2   \over \alpha a_2}
\right)^2 +
\left( 
{ 2 \pi m_3   \over \alpha a_3}
\right)^2
\right]
 +
J
\right)
\times
$$
$$
\times
\sum_{ n_1 + n_2 + n_3 = J }
{1 \over  {n_1} ! }
{1 \over  {n_2} ! }
{1 \over  {n_3} ! }
\left(
{2 \pi i m_1  \over \alpha a_1}
\right)^{{{2n}}_1}
\left(
{2 \pi i m_2  \over \alpha a_2}
\right)^{{{2n}}_2}
\left(
{2 \pi i m_3  \over \alpha a_3}
\right)^{{{2n}}_3}
 .
$$
\par
The last sum according to Newton's binomial theorem is
$$
S_J
=
\left(
{\alpha^2  \over k }\right)\ 
\left( {\alpha \over 2 \pi k} \right)^{3/2}
\sqrt {{2 \pi k  \over a_1 }}
\sqrt {{2 \pi k  \over a_2 }}
\sqrt {{2 \pi k  \over a_3 }}
\times
\eqno (32)$$
$$
\times
\exp \left[
{k \over \alpha }\left(
{2 \pi m_1  \over \alpha a_1}
\right)^2
\right]\ 
\exp \left[
{k \over \alpha }\left(
{2 \pi m_2  \over \alpha a_2}
\right)^2
\right]\ 
\exp \left[
{k \over \alpha }\left(
{2 \pi m_3  \over \alpha a_3}
\right)^2
\right]\ 
\times
$$
$$
\times
\left(
{k \over \alpha} 
\left[
\left( 
{ 2 \pi m_1   \over \alpha a_1}
\right)^2 + 
\left( 
{ 2 \pi m_2   \over \alpha a_2}
\right)^2 +
\left( 
{ 2 \pi m_3   \over \alpha a_3}
\right)^2
\right]
 +
J
\right)
\times
$$
$$
\times
{1 \over J }!
\left[
\left(
{k \over \alpha }\right)\ 
\left(
\left( 
{ 2 \pi i m_1   \over \alpha a_1}
\right)^2 + 
\left( 
{ 2 \pi i m_2   \over \alpha a_2}
\right)^2 +
\left( 
{ 2 \pi i m_3   \over \alpha a_3}
\right)^2
\right)
\right]^J
 .
$$
\par
We see, that result $S_J$ depends on two variables $J = n_1 + n_2 + n_3$ and $M$
$$
M = 
\left(
{k \over \alpha }\right)\ 
\left(
\left( 
{ 2 \pi m_1   \over \alpha a_1}
\right)^2 + 
\left( 
{ 2 \pi m_2   \over \alpha a_2}
\right)^2 +
\left( 
{ 2 \pi m_3   \over \alpha a_3}
\right)^2
\right) .
\eqno (33)$$
\par
With this variables (32) reads
$$
S_J
=
\left(
{\alpha^2  \over k }\right)\ 
\left( {\alpha \over 2 \pi k} \right)^{3/2}
\sqrt {{2 \pi k  \over a_1 }}
\sqrt {{2 \pi k  \over a_2 }}
\sqrt {{2 \pi k  \over a_3 }}
\times
\eqno (34)$$
$$
\times\ 
e^M
\left(
M + J
\right)
{1 \over J }!
(-M)^J
 .
$$
\par
The last effort is to calculate sum on $J$
$$
\sum_{J=0}^{\infty}
S_J =
\left(
{\alpha^2  \over k }\right)\ 
\left( {\alpha \over 2 \pi k} \right)^{3/2}
\sqrt {{2 \pi k  \over a_1 }}
\sqrt {{2 \pi k  \over a_2 }}
\sqrt {{2 \pi k  \over a_3 }}
\times
\eqno (35)$$
$$
\times\ 
e^M
\sum_{J=0}^{\infty}
\left(
M + J
\right)
{1 \over J }!
(-M)^J ;
$$
$$
\sum_{J=0}^{\infty}
\left(
M + J
\right)
{1 \over J }!
(-M)^J =
M e^{-M} +
(-M) e^{-M} =
0 .
\eqno (36)$$
\par\noindent
that is coefficient by $P_{ m_1 m_2 m_3 } (t)$ in (29) is zero and (29) reads
$$
\sum_{ n_1 = 0 }^{ \infty }
\sum_{ n_2 = 0 }^{ \infty }
\sum_{ n_3 = 0 }^{ \infty }
c_{{m}_1 n_1}
c_{{m}_2 n_2}
c_{{m}_3 n_3}
( n_1 + n_2 + n_3 )
A_{ m_1 m_2 m_3 n_1 n_2 n_3 } (t)
= 0 .
\eqno (37)$$
\par
This result is a little disappointment, because we still not get desired equation for $P_{ m_1 m_2 m_3 } (t)$. We need insistence to achieve success. The result is already near.
\par
Differentiate (37) again  on $t$
$$
\sum_{ n_1 = 0 }^{ \infty }
\sum_{ n_2 = 0 }^{ \infty }
\sum_{ n_3 = 0 }^{ \infty }
c_{{m}_1 n_1}
c_{{m}_2 n_2}
c_{{m}_3 n_3}
( n_1 + n_2 + n_3 )
A'_{ m_1 m_2 m_3 n_1 n_2 n_3 } (t)
= 0 .
\eqno (38)$$
\par\noindent
and substitute $A'$ from (22)
$$
\sum_{ n_1 = 0 }^{ \infty }
\sum_{ n_2 = 0 }^{ \infty }
\sum_{ n_3 = 0 }^{ \infty }
c_{{m}_1 n_1}
c_{{m}_2 n_2}
c_{{m}_3 n_3}
( n_1 + n_2 + n_3 )
\sum_{{j=1}}^{{j=3}}
\left[
\alpha n_j +
k\  \left(
{2 \pi m_j  \over \alpha a_j }
\right)^2
\right]\ 
A_{ m_1 m_2 m_3 n_1 n_2 n_3 } (t)
=
\eqno (39)$$
$$
=
2 \pi i\ 
\left(
{\alpha \over k }\right)\ 
\left( {\alpha \over 2 \pi k} \right)^{3/2} 
\sum_{ n_1 = 0 }^{ \infty }
\sum_{ n_2 = 0 }^{ \infty }
\sum_{ n_3 = 0 }^{ \infty }
c_{{m}_1 n_1}
c_{{m}_2 n_2}
c_{{m}_3 n_3}
( n_1 + n_2 + n_3 )
\times
$$
$$
\times
\left(
{m_1 \over a_1 }b_{{m}_1 n_1}
a_{{m}_2 n_2}
a_{{m}_3 n_3} +
{m_2 \over a_2 }a_{{m}_1 n_1}
b_{{m}_2 n_2}
a_{{m}_3 n_3} +
\right.
$$
$$
+
\left.
{m_3 \over a_3 }a_{{m}_1 n_1}
a_{{m}_2 n_2}
b_{{m}_3 n_3}
\right)\ 
P_{ m_1 m_2 m_3 } (t)\  
 .
$$
\par
Let us use (37) to remove terms with $\left(
{2 \pi m_j  \over \alpha a_j }
\right)^2$ in (39). To calculate coefficient before $P_{ m_1 m_2 m_3 } (t)$ we perform once again summation on group of terms with constant $(n_1 + n_2 + n_3 ) = J$. The sum for each group is calculated as before, but this time this sum is multiplied by $J$ before final summation on $J$
$$
\sum_{ n_1 = 0 }^{ \infty }
\sum_{ n_2 = 0 }^{ \infty }
\sum_{ n_3 = 0 }^{ \infty }
c_{{m}_1 n_1}
c_{{m}_2 n_2}
c_{{m}_3 n_3}
\alpha
( n_1 + n_2 + n_3 )^2
A_{ m_1 m_2 m_3 n_1 n_2 n_3 } (t)
=
\eqno (40)$$
$$
=
\left(
{\alpha^2  \over k }\right)\ 
\left( {\alpha \over 2 \pi k} \right)^{3/2}
\sqrt {{2 \pi k  \over a_1 }}
\sqrt {{2 \pi k  \over a_2 }}
\sqrt {{2 \pi k  \over a_3 }}
e^M
\sum_{J=0}^{\infty}
\left(
M + J
\right)
{J \over J }!
(- M)^J
P_{ m_1 m_2 m_3 } (t)\  
 .
$$
\par
This time the sum on $J$ is not zero
$$
\sum_{J=0}^{\infty}
\left(
M + J
\right)
{J \over J }!
(-M)^J =
- M^2 e^{-M} +
(-M)^2 e^{-M} -
M e^{-M} =
-M e^{-M} ,
\eqno (41)$$
\par\noindent
and we get finally equation for pressure, which follows from incompressibility condition

\boxit{
$$
\sum_{ n_1 = 0 }^{ \infty }
\sum_{ n_2 = 0 }^{ \infty }
\sum_{ n_3 = 0 }^{ \infty }
c_{{m}_1 n_1}
c_{{m}_2 n_2}
c_{{m}_3 n_3}
( n_1 + n_2 + n_3 )^2
A_{ m_1 m_2 m_3 n_1 n_2 n_3 } (t)
=
\eqno (42)$$
$$
=
-
\left(
{\alpha \over k }\right)\ 
\left( {\alpha \over 2 \pi k} \right)^{3/2}
\sqrt {{2 \pi k  \over a_1 }}
\sqrt {{2 \pi k  \over a_2 }}
\sqrt {{2 \pi k  \over a_3 }}
M
P_{ m_1 m_2 m_3 } (t)\  
 .
$$
}

\par
(42) is algebraic equation, but in the same time it is Fourier transform of some differential equation for original unknown variable $n$. Namely according to (33) $M = \left(
{k \over \alpha }\right)\ 
\left(
\left( 
{ 2 \pi m_1   \over \alpha a_1}
\right)^2 + 
\left( 
{ 2 \pi m_2   \over \alpha a_2}
\right)^2 +
\left( 
{ 2 \pi m_3   \over \alpha a_3}
\right)^2
\right)$ . Each $m_i^2$ term in (42) is Fourier transform of second partial derivative of $p$ on corresponding space variable, their sum is Fourier transform of Laplace operator. Therefore (42) is Fourier transform of Poisson's equation for pressure.
\par
Solve this equation for $P_{ m_1 m_2 m_3 } (t)$ and get
$$
P_{ m_1 m_2 m_3 } (t) =
{-1 \over M }\left(
{k \over \alpha }\right)\ 
\left( {2 \pi k  \over \alpha }\right)^{3/2}
\sqrt { {a_1  \over 2 \pi k}}
\sqrt { {a_2  \over 2 \pi k}}
\sqrt { {a_3  \over 2 \pi k}}
\times
\eqno (43)$$
$$
\times
\sum_{ n_1 = 0 }^{ \infty }
\sum_{ n_2 = 0 }^{ \infty }
\sum_{ n_3 = 0 }^{ \infty }
c_{{m}_1 n_1}
c_{{m}_2 n_2}
c_{{m}_3 n_3}
( n_1 + n_2 + n_3 )^2
A_{ m_1 m_2 m_3 n_1 n_2 n_3 } (t)
 .
$$
\par
Let us take values of $c_{mn}$ from (25)
$$
P_{ m_1 m_2 m_3 } (t) =
{-\exp (- M / 2 )  \over M }\left(
{k \over \alpha }\right)\ 
\left( {2 \pi k  \over \alpha }\right)^{3/2}
\times
\eqno (44)$$
$$
\times
\sum_{ \nu_1 = 0 }^{ \infty }
\sum_{ \nu_2 = 0 }^{ \infty }
\sum_{ \nu_3 = 0 }^{ \infty }
\left(
{2k \over \alpha }\right)^{ ( \nu_1 + \nu_2 + \nu_3 ) /2}
\left(
{2 \pi i m_1  \over \alpha a_1}
\right)^{{\nu}_1}
\left(
{2 \pi i m_2  \over \alpha a_2}
\right)^{{\nu}_2}
\left(
{2 \pi i m_3  \over \alpha a_3}
\right)^{{\nu}_3}
( \nu_1 + \nu_2 + \nu_3 )^2
A_{ m_1 m_2 m_3 \nu_1 \nu_2 \nu_3 } (t)
 .
$$
\par\noindent
and substitute result to (23)
$$
{d \over dt }A_{ m_1 m_2 m_3 n_1 n_2 n_3 } (t)
+ 
\sum_{{j=1}}^{{j=3}}
\left[
\alpha n_j +
k\  \left(
{2 \pi m_j  \over \alpha a_j }
\right)^2
\right]\ 
A_{ m_1 m_2 m_3 n_1 n_2 n_3 } (t)
=
\eqno (45)$$
$$
=
2 \pi i\ 
a_{m_1 n_1}
a_{m_2 n_2}
a_{m_3 n_3}\ 
\left(
{m_1 \over a_1 } {b_{m_1 n_1} \over a_{ m_1 n_1}} +
{m_2 \over a_2 } {b_{m_2 n_2} \over a_{ m_2 n_2}} +
{m_3 \over a_3 } {b_{m_3 n_3} \over a_{ m_3 n_3}}
\right)\ 
{ {\exp (- M / 2 )} \over { (-M) }}
\times
$$
$$
\times
\sum_{ \nu_1 = 0 }^{ \infty }
\sum_{ \nu_2 = 0 }^{ \infty }
\sum_{ \nu_3 = 0 }^{ \infty }
\left(
{2k \over \alpha }\right)^{ ( \nu_1 + \nu_2 + \nu_3 ) /2}
\left(
{2 \pi i m_1  \over \alpha a_1}
\right)^{{\nu}_1}
\left(
{2 \pi i m_2  \over \alpha a_2}
\right)^{{\nu}_2}
\left(
{2 \pi i m_3  \over \alpha a_3}
\right)^{{\nu}_3}
( \nu_1 + \nu_2 + \nu_3 )^2
A_{ m_1 m_2 m_3 \nu_1 \nu_2 \nu_3 } (t)
 .
$$
\par
To get final result, substitute values of $a_{mn}$ from (17)

\boxit{
$$
{d \over dt }A_{ m_1 m_2 m_3 n_1 n_2 n_3 } (t)
+ 
\sum_{{j=1}}^{{j=3}}
\left[
\alpha n_j +
k\  \left(
{2 \pi m_j  \over \alpha a_j }
\right)^2
\right]\ 
A_{ m_1 m_2 m_3 n_1 n_2 n_3 } (t)
=
\eqno (46)$$
$$
=
{\exp ( M )  \over (-M)}
{1 \over  2^{{n}_1 + n_2 + n_3} n_1 ! n_2 ! n_3 !}
\left(
{-2k \over \alpha }\right)^{{(} n_1 + n_2 + n_3 ) /2}
\times
$$
$$
\times\ 
\left(
{2 \pi m_1  \over \alpha a}
\right)^{{n}_1}
\left(
{2 \pi m_2  \over \alpha a}
\right)^{{n}_2}
\left(
{2 \pi m_3  \over \alpha a}
\right)^{{n}_3}
\sum_{{j=1}}^{{j=3}}
\left[
\alpha n_j +
k\  \left(
{2 \pi m_j  \over \alpha a_j }
\right)^2
\right]\ 
\times
$$
$$
\times
\sum_{ \nu_1 = 0 }^{ \infty }
\sum_{ \nu_2 = 0 }^{ \infty }
\sum_{ \nu_3 = 0 }^{ \infty }
\left(
{-2k \over \alpha }\right)^{ ( \nu_1 + \nu_2 + \nu_3 ) /2}
\left(
{2 \pi m_1  \over \alpha a_1}
\right)^{{\nu}_1}
\left(
{2 \pi m_2  \over \alpha a_2}
\right)^{{\nu}_2}
\left(
{2 \pi m_3  \over \alpha a_3}
\right)^{{\nu}_3}
( \nu_1 + \nu_2 + \nu_3 )^2
A_{ m_1 m_2 m_3 \nu_1 \nu_2 \nu_3 } (t)
 .
$$
}
\par
We totally eliminated pressure from dynamic equation. Thus the problem is reduced to the system of ordinary linear differential equations (46). Such systems are solved by Euler's exponential substitution. The only difficulty is, that the number of variables is indefinitely great. We try to integrate (46) using special properties of its matrix.
\section{Special eigenvalue problem}
\par
In this section we consider purely algebraic eigenvalue problem for special form of matrices.
\par
Let the matrix have the following form
$$
M_{ij} = D_{ij} + P_{ij} ;
\eqno (47)$$
\par\noindent
where $D_{ij}$ - diagonal matrix
$$
D_{ij} = \left\{ \matrix { 0;\ \ i \ne j \cr d_{i;}\ \ i = j} \right.
\eqno (48)$$
\par\noindent
$P_{ij}$ - diadic product matrix
$$
P_{ij} = l_i r_j .
\eqno (49)$$
\par
The eigenvalue problem consists of following tasks:
\par\noindent
1) To find a set of eigenvectors $x$ such, that matrix product of $M$ and $x$ is proportional to $x$
$$
M_{ij} x_i = \lambda x_j .
\eqno (50)$$
\par\noindent
2) To find corresponding set of proportionality coefficients - eigenvalues $\lambda$.
\par
For our special form of matrix
$$
( D_{ij} - \lambda\  \delta_{ij} ) x_j +
l_i \  ( r_k x_k ) = 0;
\eqno (51)$$
\par
Let us denote
$$
S = r_k x_k ;
\eqno (52)$$
\par
Then
$$
( d_1 - \lambda )\ {x_1 \over l_1 }=
( d_2 - \lambda )\ {x_2 \over l_2 }=
... = - S .
\eqno (53)$$
\par
This gives very simple expression for components of eigenvector, provided eigenvalue $\lambda$ is evident
$$
x_i = {-S l_i   \over ( d_i - \lambda )} .
\eqno (54)$$
\par
Substitute this expression to (52) and get
$$
\sum_k 
{-S\  l_k r_k   \over ( d_k - \lambda )} = S .
\eqno (55)$$
\par
Two cases are possible here. The common case is $S != 0$.
$$
\sum_k 
{l_k r_k  \over ( d_k - \lambda )} + 1 = 0.
\eqno (56)$$
\par
(56) gives equation for $\lambda$. When all $d_i $ are different, algebraic equation (56) has degree $n$, where $n$ - dimension of matrix $M$. 
\par
If all roots of (56) are different, we get the full set of eigenvalues, then from (54) we find full set of eigenvectors. Exact value of $S$ is of no meaning, we can put for example $S=1$ in (54).
\par
For example, when all $( l_k r_k )$ are positive or all are negative and all $d_i$ are real (and different - see above), one could guarantee that all $n$ roots of (56) are real and different. This follows from the fact, that $d_i$ separate the roots of (56). Therefore there are $n-1$ roots between $d_i$ and one more root $x_1  min (d_i )$ (the case $l_i r_i  0$). This fact makes numeric evaluation of roots rather simple. Unfortunately we deal with quite opposite case - signs of our $( l_k r_k )$ are alternating. Nevertheless we shall find roots - see below.
\par
When some $d_i$ are equal, equation (56) has less roots then matrix degree. We can consider this case as confluent. In this case additional eigenvalues $\lambda$ (besides roots of (56)) must be equal to iterated value $d_i$ and we must put for these eigenvalues $S=0$ in (53). Then all $x_i $ , besides that in columns, corresponding to iterated $d_i$, must be zero. The rest nonzero $x_i$ must satisfy orthogonality condition (52) (with $S=0$).
\par
Let us consider the eigenvalue problem for transposed matrix $M^T $ . Let us denote $y$ - eigenvectors for $M^T$. Then
$$
( D_{ij} - \lambda\  \delta_{ij} ) y_i +
( l_k y_k ) r_j = 0;
\eqno (57)$$
\par\noindent
or
$$
( a_i - \lambda ) {y_i  \over r_i} = -  S .
\eqno (58)$$
$$
y_i = {-S r_i  \over ( d_i - \lambda )} .
\eqno (59)$$
$$
S = l_k y_k =
\sum_k {-S\  l_k r_k  \over ( d_k - \lambda )}
.
\eqno (60)$$
\par
We get for $\lambda$ equation (56) once again - eigenvalues of conjugated problems are equal. Components of eigenvectors for conjugated problem are calculated from (59), where $S$ is arbitrary nonzero number, for example $S = 1$.
\par
Eigenvector for conjugated problems with different eigenvalues $\lambda$ and $\mu$ are orthogonal:
$$
\sum_k x_k y_k =
\sum_k
{- l_k  \over ( d_k - \lambda )}
{- r_k  \over ( d_k - \mu )} =
\sum_k
{l_k r_k   \over \lambda - \mu}
\left(
{1 \over ( d_k - \lambda )} -
{1 \over ( d_k - \mu )}
\right) =
\eqno (61)$$
$$
=
{1 \over \lambda - \mu}
\left(
\sum_k
{l_k r_k   \over ( d_k - \lambda )}
- 
\sum_k
{l_k r_k   \over ( d_k - \mu )}
\right) 
=
{1 \over \lambda - \mu}
\left(
-1 + 1
\right) = 0.
$$
\section{Application to equation (46)}
\par
Let us return to our equation (46). Comparing with previous section, we could identify matrix components in the following way.
\par
Diagonal components of the matrix are equal to:
$$
d_{{n}_1 n_2 n_3} =
\sum_{{j=1}}^{{j=3}}
\left[
\alpha n_j +
k\  \left(
{2 \pi m_j  \over \alpha a_j }
\right)^2
\right] .
\eqno (62)$$
\par
Diagonal components depend only on $M$ (see (33)) and $J = \sum n_l$. Therefore we deal with confluent case - see previous section. Some eigenvalues are equal to $d_{{n}_1 n_2 n_3}$. 
\par
Nondiagonal components of the matrix are equal to:
$$
l_{{n}_1 n_2 n_3} =
{\exp ( M ) \over M } 
{1 \over {2^{n_1 + n_2 + n_3} n_1 ! n_2 ! n_3 !} }
\left(
{-2k \over \alpha }\right)^{{(} n_1 + n_2 + n_3 ) /2}
\times
\eqno (63)$$
$$
\times\ 
\left(
{2 \pi m_1  \over \alpha a}
\right)^{{n}_1}
\left(
{2 \pi m_2  \over \alpha a}
\right)^{{n}_2}
\left(
{2 \pi m_3  \over \alpha a}
\right)^{{n}_3}
\sum_{{j=1}}^{{j=3}}
\left[
\alpha n_j +
k\  \left(
{2 \pi m_j  \over \alpha a_j }
\right)^2
\right]\ 
 .
$$
$$
r_{{\nu}_1 \nu_2 \nu_3} =
\left(
{-2k \over \alpha }\right)^{ ( \nu_1 + \nu_2 + \nu_3 ) /2}
\left(
{2 \pi m_1  \over \alpha a_1}
\right)^{{\nu}_1}
\left(
{2 \pi m_2  \over \alpha a_2}
\right)^{{\nu}_2}
\left(
{2 \pi m_3  \over \alpha a_3}
\right)^{{\nu}_3}
( \nu_1 + \nu_2 + \nu_3 )^2
 .
\eqno (64)$$
\par
In all cases we replaced index $(i)$ with multiindex ${( n_1 n_2 n_3 )}.$
\par
Characteristic equation for $\lambda$ is
$$
\sum_{ n_1 = 0 }^{ \infty }
\sum_{ n_2 = 0 }^{ \infty }
\sum_{ n_3 = 0 }^{ \infty }
{l_{{n}_1 n_2 n_3} r_{{n}_1 n_2 n_3} 
\over \lambda -
{d_{{n}_1 n_2 n_3} }}
=
\eqno (65)$$
$$
=
{e^M  \over M }\sum_{ n_1 = 0 }^{ \infty }
\sum_{ n_2 = 0 }^{ \infty }
\sum_{ n_3 = 0 }^{ \infty }
{ \sum_{{j=1}}^{{j=3}}
\left[
\alpha n_j +
k\  \left(
{2 \pi m_j  \over \alpha a_j }
\right)^2
\right]\  
\over  \lambda - \sum_{{j=1}}^{{j=3}}
\left[
\alpha n_j +
k\  \left(
{2 \pi m_j  \over \alpha a_j }
\right)^2
\right]\ }
\times
$$
$$
\times\ 
\left(
{-k \over \alpha }\right)^{ ( n_1 + n_2 + n_3 )}
\left(
{2 \pi m_1  \over \alpha a_1}
\right)^{{2n}_1}
\left(
{2 \pi m_2  \over \alpha a_2}
\right)^{{2n}_2}
\left(
{2 \pi m_3  \over \alpha a_3}
\right)^{{2n}_3}
{( n_1 + n_2 + n_3 )^2 
\over  n_1 ! n_2 ! n_3 !} =
1 .
$$
\par
Let us once again calculate partial sum $S_J$ of terms, for which $n_1 + n_2 + n_3 = J$. We have
$$
S_J =
{e^M  \over M }\left(
{\alpha (J + M) J^2  \over \lambda - \alpha (J + M) }
\right)
{(-M)^J  \over J }! .
\eqno (66)$$
$$
\sum_{J=0}^{\infty}
S_J =
\left(
1 - M
\right)
- {\lambda \over \alpha }{e^M  \over M }M^{{-(} M - \lambda / \alpha )}\ 
\times
\eqno (67)$$
$$
\times\ 
\left[
\gamma ( ( 2 + M - \lambda / \alpha ), M) -
\gamma ( ( 1 + M - \lambda / \alpha ), M)
\right]
= 1
.
$$
\par
where $\gamma ( p, z)$ is incomplete gamma - function (see [4], [5])
$$
\gamma ( p, z) =
\int_{t=0}^{t=z}
e^{-t} t^{p-1}
dt .
\eqno (68)$$
\par
(65) reads now
$$
M
+ {\lambda \over \alpha }{e^M  \over M }M^{{-(} M - \lambda / \alpha )}\ 
\left[
\gamma ( ( 2 + M - \lambda / \alpha ), M) -
\gamma ( ( 1 + M - \lambda / \alpha ), M)
\right]
= 0
.
\eqno (69)$$
\par
We see, that all eigenvalues $\lambda$ are proportional to $\alpha$
$$
\lambda = \xi \alpha ,
\eqno (70)$$
\par
where coefficients $\xi$ are roots of equation

\boxit{
$$
M + \xi
{e^M  \over M }M^{{-(} M - \xi )}\ 
\left[
\gamma ( ( 2 + M - \xi ), M) -
\gamma ( ( 1 + M - \xi ), M)
\right]
= 0
.
\eqno (71)$$
}

\par
We can further simplify (71) using known recurrence (see [4], 9.2)
$$
\gamma (p+1, z) =
p \gamma (p, z) - x^p e^{-x} .
\eqno (72)$$
$$
M + \xi
{e^M  \over M } M^{- ( M - \xi )}\ 
\left[
( 1 + M - \xi ) \gamma ( ( 1 + M - \xi ), M) 
\right.
-
\eqno (73)$$
$$
\left.
-
M^{( 1 + M - \xi )} e^{-M} -
\gamma ( ( 1 + M - \xi ), M)
\right]
= 0
.
$$
$$
M + \xi
{e^M  \over M }M^{{-(} M - \xi )}\ 
\left[
( M - \xi ) \gamma ( ( 1 + M - \xi ), M) -
M^{{(} 1 + M - \xi )} e^{-M}
\right]
= 0
.
\eqno (74)$$
$$
M - \xi + \xi
{e^M  \over M }M^{{-(} M - \xi )}\ 
( M - \xi ) \gamma ( ( 1 + M - \xi ), M)
= 0
.
\eqno (75)$$
\par
One exact root of equation (75) we can find easily - this is root 
$$
\xi = M .
\eqno (76)$$
\par
Another roots satisfy reduced equation

\boxit{
$$
1 + \xi
{e^M  \over M }M^{{-(} M - \xi )}\ 
\gamma ( ( 1 + M - \xi ), M)
= 0
.
\eqno (77)$$
}
\par
Another form of (77) we obtain using modified incomplete gamma function
$$
\gamma^* (p, x) =
{x^{-p}  \over { \Gamma (p) } }
\gamma (p, x) .
\eqno (78)$$
\par
This form is
$$
1 + \xi
{e^M} 
\Gamma ( 1 + M - \xi )\ 
\gamma^* ( ( 1 + M - \xi ), M)
= 0
.
\eqno (79)$$
\par
Advantage of this form is that $\gamma^* ( p, x)$ is a single valued analytic function of $p$ and $x$ possessing no finite singularities.
\par
We calculate some roots of equation (77) using definition of incomplete gamma-function (66 - 67). Namely we keep only finite number of terms in the sum (66 - 67) and solve resulting algebraic equation using Newton's method. As the number of terms increase, the roots converge rapidly.
\par
As initial approximation for roots we use position of poles, that is we search roots in the close vicinity of the pole. We keep corresponding term and approximate contribution of the rest poles by two first terms of Taylor series. So we have quadratic equation. When this equation has two conjugated roots, Newton's iterations converge to two conjugated roots of full equation. When quadratic equation has two real roots, one root is really located in the poles vicinity and another is far enough, so that Newton's iterations diverge.
\par
The results of our calculations are presented in the Table 1 (see APPENDIX). We see, that:
\par\noindent
- All roots have positive real part. Therefore exists root with the least real part and roots can be ordered according their real part in ascending order. In the following we suppose that such ordering is done.
\par\noindent
- For each $M$ only finite number of roots are complex and imaginary part of roots decreases with root number.
\par\noindent
- Starting from some root all roots are real (tail).
\par\noindent
- Real roots are located very close to poles (natural numbers) and rapidly get indistinguishable.
\par
On this stage we shall content ourself by these experimental results and shall not try to give them rigorous proof. The theory of distribution of roots of analytic functions in question is rather ample (see [6], [7], [8]).
\par
As we see from previous section, besides roots of (77) there exist eigenvalues, which are exactly equal to diagonal values $d_{{n}_1 n_2 n_3}$ (62). They correspond to zero-pressure solutions from our work [2]. Orthogonality conditions (52), which must be satisfied by eigenfunctions, according to (64) and (43) mean, that $P_{{m}_1 m_2 m_3} (t) = 0$. 
\section{Solution of Cauchy problem}
\par
In this section we describe construction of solution of linearized Fokker - Planck equation for incompressible fluid.
\par
1) To setup Cauchy problem we must set initial value of $n$ variable. This initial value $n_0 = n_0 (x_1 , x_2 , x_3 , v_1 , v_2 , v_3 )$ must satisfy incompressibility condition (1). There is no need to set initial value of pressure, because it is fully determined by $n$ (see (42)).
\par
2) Calculate Fourier coefficients $A_{ m_1 m_2 m_3 p_1 p_2 p_3 } (0)$ (see [3] for details):

$$
A_{ m_1 m_2 m_3 p_1 p_2 p_3 }(0) =
{1 \over 2^{{p}_1 + p_2 + p_3}
p_1 ! 
p_2 !
p_3 !}\ 
\left( 
{\alpha \over 2 \pi k} 
\right)^{ {3 \over 2 }}
{1 \over a_1 a_2 a_3}
\times
\eqno (80)$$
$$
\times
\exp
\left[
 {\alpha \over 2k }\ 
\left(
{4 \pi k  \over \alpha^2}
\right)^2
\left(
\left(
{m_1  \over a_1}
\right)^2 +
\left(
{m_2  \over a_2}
\right)^2 +
\left(
{m_3  \over a_3}
\right)^2
\right)
\right] 
\times
$$
$$
\times
\int_0^{{a}_1}
dx_1
\int_0^{{a}_2}
dx_2
\int_0^{{a}_3}
dx_3
\int_{{-} \infty}^{\infty}
\int_{{-} \infty}^{\infty}
\int_{{-} \infty}^{\infty}
n_0 (x_1 , x_2 , x_3 , v_1 , v_2 , v_3 )\ \  
\psi_{ m_1 m_2 m_3 p_1 p_2 p_3 }
dv_1
dv_2
dv_3
.
$$
\par\noindent
where
$$
\psi_{ m_1 m_2 m_3 n_1 n_2 n_3 } =
\prod_{{j=1}}^{{j=3}}
\exp 
\left( 
- 2 \pi i {m_j \over a }( x_j + {v_j \over \alpha }) 
\right) \ 
H_{{n}_j}
\left(
\sqrt {{\alpha \over 2k }}
\left(
v_j +
{4 \pi i {m_j} k  \over \alpha^2 a_j}
\right)
\right)
 .
\eqno (81)$$
\par
3) Change from eigenfunctions of simple Fokker - Planck equations to eigenfunctions of linearized Fokker - Planck equation
for incompressible fluid. According to (61) projection of vector $A_k$ on eigenvector $x_{\mu}$, corresponding to eigenvalue $\lambda$, is
$$
A_{\lambda} = 
{ \sum_k A_k y_k   \over  \sum_k y_k y_k }
 =
\left(
{ \sum_k A_k {r_k \over ( \lambda - d_k )} }
\right)\ \ 
 /\ \ 
\left(
{ \sum_k {r_k^2  \over ( \lambda - d_k )^2} }
\right)
\eqno (82)$$
\par\noindent
or according to (64)
$$
A_{ m_1 m_2 m_3 \lambda }(0)
 =
\left(
 \sum_{{\nu}_1 \nu_2 \nu_3} 
A_{ m_1 m_2 m_3 \nu_1 \nu_2 \nu_3 }(0)\ 
\left(
{-2k \over \alpha }\right)^{ ( \nu_1 + \nu_2 + \nu_3 ) /2}
\right.
\times
\eqno (83)$$
$$
\times
\left(
{2 \pi m_1  \over \alpha a_1}
\right)^{{\nu}_1}
\left(
{2 \pi m_2  \over \alpha a_2}
\right)^{{\nu}_2}
\left(
{2 \pi m_3  \over \alpha a_3}
\right)^{{\nu}_3}
\times
$$
$$
\times
\left.
{ ( \nu_1 + \nu_2 + \nu_3 )^2  
 \over ( \lambda - d_{{\nu}_1 \nu_2 \nu_3} )} 
\right)\ \ 
 /\ \ 
\left(
{ \sum_k {r_{{n}_1 n_2 n_3}^2  \over ( \lambda - d_{{\nu}_1 \nu_2 \nu_3} )^2} }
\right)
$$
\par\noindent
where $d_{{n}_1 n_2 n_3}$ is defined by (62), $r_{{n}_1 n_2 n_3}$ is defined by (64).
\par
4) Initial field can contain some zero pressure solutions. Let us suppose, that there exist a group of coefficients with constant $J = \sum \nu_l$, for which
$$
 \sum_{ \nu_1 + \nu_2 + \nu_3 = J} 
A_{ m_1 m_2 m_3 \nu_1 \nu_2 \nu_3 }(0)\ 
\left(
{2 \pi m_1  \over \alpha a_1}
\right)^{{\nu}_1}
\left(
{2 \pi m_2  \over \alpha a_2}
\right)^{{\nu}_2}
\left(
{2 \pi m_3  \over \alpha a_3}
\right)^{{\nu}_3}
= 0 .
\eqno (84)$$
\par
This group does not contribute to any $A_{ m_1 m_2 m_3 \lambda }$ (see 83). Such groups, if present, we must consider separately. They correspond to zero pressure solutions of our work [2].
\par
5) Given initial values of $A_{\lambda}$ we can calculate their values for the arbitrary moment $t$ according to exponential law
$$
A_{ m_1 m_2 m_3 \lambda }(t) =
e^{{-} \lambda t }
A_{ m_1 m_2 m_3 \lambda }(0) .
\eqno (85)$$
\par
6) Evolution of zero pressure solution is determined by exponential multiplier $e^{{-} \alpha (J+M) t}$.
\par
7) Inverse transition from $A_{\lambda}$ to $A_k$ is 
$$
A_k = \sum_{\lambda}
A_{\lambda} {l_k  \over ( \lambda - d_k )}
\eqno (86)$$
\par\noindent
or according to (63)
$$
A_{ m_1 m_2 m_3 n_1 n_2 n_3 }(t) 
= \sum_{\lambda}
A_{ m_1 m_2 m_3 \lambda }(t)
{\exp ( M )  \over M }\times
\eqno (87)$$
$$
\times
{1 \over  2^{{n}_1 + n_2 + n_3} n_1 ! n_2 ! n_3 !}
\left(
{-2k \over \alpha }\right)^{{(} n_1 + n_2 + n_3 ) /2}
\times
$$
$$
\times\ 
\left(
{2 \pi m_1  \over \alpha a}
\right)^{{n}_1}
\left(
{2 \pi m_2  \over \alpha a}
\right)^{{n}_2}
\left(
{2 \pi m_3  \over \alpha a}
\right)^{{n}_3}
\sum_{{j=1}}^{{j=3}}
\left[
\alpha n_j +
k\  \left(
{2 \pi m_j  \over \alpha a_j }
\right)^2
\right]\ 
{1
\over ( \lambda - d_{{n}_1 n_2 n_3} )} .
$$
\par\noindent
$d_{{n}_1 n_2 n_3}$ defined by (62).
\par
8) Contribution from zero pressure solutions must be added to (87).
\par
9) Pressure for each moment of time is determined by (44).
\shead{DISCUSSION}
\par\noindent
We see that spectral properties of Fokker - Planck linearized differential operator for incompressible fluid are different from properties of usual operator. General spectrum structure is roughly similar, but nearest to zero eigenvalues are complex. Therefore most slowly damping modes are most strongly vibrating - very interesting result. Generally all modes are damping with time, flows tend to the rest.

\rule{2in}{1pt}
\shead{REFERENCES}

\begin{IPlist}
\IPitem{{[1]}}
Igor A. Tanski. Fokker - Planck equation for incompressible fluid. 
arXiv:0812.2303v2 [nlin.CD] 3 Feb 2009
\IPitem{{[2]}}
Igor A. Tanski. Two simple solutions of nonlinear Fokker - Planck equation for incompressible fluid. 
arXiv:0812.4795v2 [nlin.CD] 25 Feb 2009
\IPitem{{[3]}}
Igor A. Tanski. Spectral decomposition of 3D Fokker - Planck differential operator. 
arXiv:nlin/0607050v3 [nlin.CD] 25 Jun 2007
\IPitem{{[4]}}
H. Bateman, A. Erdelyi. Higher transcendental functions. 
vol. 2, Mc Graw-Hill, New York, 1953
\IPitem{{[5]}}
M. Abramovitz, I. A. Stegun, Handbook of Mathematical Functions. 
National Bureau of Standards, 1970
\IPitem{{[6]}}
K. S. Koelbig. On the zeros of incomplete gamma function. 
Mathematics of Computation, vol. 26, num. 119,  Jul 1972
\IPitem{{[7]}}
Walter Gautschi. The incomplete gamma functions since Tricomi. 
In Tricomi's Ideas and Contemporary Applied Mathematics, Atti dei Convegni Lincei, n. 147, Accademia Nazionale dei Lincei
\IPitem{{[8]}}
A. M. Sedlecki. Zeros of Mittag-Leffler functions. 
Matematicheskie zametki, vol. 68, num. 5, Nov. 2000

\end{IPlist}\newpage

\par
APPENDIX 1
\par
Roots of equation (71)
\par
Root $\xi = M$ is omitted

\begin{tabular} { || l | l | l | l || } \hline \hline
NN&M=1&M=2&M=3\\ \hline \hline
1&3.84958810 - 1.92315575 i&4.52745332 - 3.27206660 i&5.04119504 - 4.34568739 i\\ \hline
2&3.84958810 + 1.92315575 i&4.52745332 + 3.27206660 i&5.04119504 + 4.34568739 i\\ \hline
3&5.94063198 - 1.14455587 i&7.06814092 - 2.67480339 i&7.90915778 - 3.92341797 i\\ \hline
4&5.94063198 + 1.14455587 i&7.06814092 + 2.67480339 i&7.90915778 + 3.92341797 i\\ \hline
5&7.69165960 - 0.29172595 i&9.14966499 - 1.97226724 i&10.24525406 - 3.33703003 i\\ \hline
6&7.69165960 + 0.29172595 i&9.14966499 + 1.97226724 i&10.24525406 + 3.33703003 i\\ \hline
7&11.000932211&11.00141191 - 1.21959567 i&12.31747194 - 2.67114557 i\\ \hline
8&11.999889388&11.00141191 + 1.21959567 i&12.31747194 + 2.67114557 i\\ \hline
9&13.000011755&12.72299911 - 0.42498830 i&14.22286668 - 1.95786294 i\\ \hline
10&13.999998866&12.72299911 + 0.42498830 i&14.22286668 + 1.95786294 i\\ \hline
11&15.000000099&14.07454792&16.00982241 - 1.21325010 i\\ \hline
12&15.999999999&14.98376860&16.00982241 + 1.21325010 i\\ \hline
13&17.000000000&16.00272440&17.72151356 - 0.43464100 i\\ \hline
14&17.999999999&16.99956146&17.72151356 + 0.43464100 i\\ \hline
15&19.000000000&18.00006506&19.98044555\\ \hline
16&19.999999999&18.99999098&19.08192627\\ \hline
17&21.000000000&20.00000116&19.98044554\\ \hline
18&22.000000000&20.99999985&21.00363177\\ \hline
19&23.000000000&22.00000001&21.99933574\\ \hline
20&24.000000000&22.99999999&23.00011373\\ \hline \hline
\end{tabular}

\begin{tabular} { || l | l | l | l || } \hline \hline
NN&M=4&M=5&M=6\\ \hline \hline
1&5.47498406 - 5.26768710 i&5.85889775 - 6.08913226 i&6.20771808 - 6.83731700 i\\ \hline
2&5.47498406 + 5.26768710 i&5.85889775 + 6.08913226 i&6.20771808 + 6.83731700 i\\ \hline
3&8.61344667 - 5.01144505 i&9.23406886 - 5.99054148 i&9.79675137 - 6.88888105 i\\ \hline
4&8.61344667 + 5.01144505 i&9.23406886 + 5.99054148 i&9.79675137 + 6.88888105 i\\ \hline
5&11.15620212 - 4.53812831 i&11.95509246 - 5.62665552 i&12.67703265 - 6.63086922 i\\ \hline
6&11.15620212 + 4.53812831 i&11.95509246 + 5.62665552 i&12.67703265 + 6.63086922 i\\ \hline
7 &13.40577133 - 3.95833350 i&14.35634208 - 5.13111612 i&15.21274497 - 6.21756080 i\\ \hline
8 &13.40577133 + 3.95833350 i&14.35634208 + 5.13111612 i&15.21274497 + 6.21756080 i\\ \hline
9 &15.47089306 - 3.31449740 i&16.55737228 - 4.55585073 i&17.53366817 - 5.70963951 i\\ \hline
10&15.47089306 + 3.31449740 i&16.55737228 + 4.55585073 i&17.53366817 + 5.70963951 i\\ \hline
11&17.40539050 - 2.62778318 i&18.61704841 - 3.92690815 i&19.70341924 - 5.13774003 i\\ \hline
12&17.40539050 + 2.62778318 i&18.61704841 + 3.92690815 i&19.70341924 + 5.13774003 i\\ \hline
13&19.24064677 - 1.91036878 i&20.56955610 - 3.25932488 i&21.75880445 - 4.51960861 i\\ \hline
14&19.24064677 + 1.91036878 i&20.56955610 + 3.25932488 i&21.75880445 + 4.51960861 i\\ \hline
15&20.99699656 - 1.17010968 i&22.43688595 - 2.56260936 i&23.72343062 - 3.86649261 i\\ \hline
16&20.99699656 + 1.17010968 i&22.43688595 + 2.56260936 i&23.72343062 + 3.86649261 i\\ \hline
17&22.70576400 - 0.39532166 i&24.23420403 - 1.84317596 i&25.61354419 - 3.18599046 i\\ \hline
18&22.70576400 + 0.39532166 i&24.23420403 + 1.84317596 i&25.61354419 + 3.18599046 i\\ \hline
19&24.07528691&25.97268205 - 1.10579718 i&27.44091387 - 2.48348667 i\\ \hline
20&24.98146932&25.97268205 + 1.10579718 i&27.44091387 + 2.48348667 i\\ \hline \hline
\end{tabular}

\begin{tabular} { || l | l | l | l || } \hline \hline
\#&M=7&M=8&M=9\\ \hline
1&6.53001156 - 7.52895368 i&6.83127967 - 8.17518985 i&7.11531330 - 8.78391538 i\\ \hline
2&6.53001156 + 7.52895368 i&6.83127967 + 8.17518985 i&7.11531330 + 8.78391538 i\\ \hline
3&10.31618762 - 7.72399088 i&10.80167947 - 8.50772212 i&11.25955129 - 9.24856515 i\\ \hline
4&10.31618762 + 7.72399088 i&10.80167947 + 8.50772212 i&11.25955129 + 9.24856515 i\\ \hline
5&13.34199931 - 7.56851364 i&13.96257983 - 8.45167276 i&14.54728640 - 9.28905550 i\\ \hline
6&13.34199931 + 7.56851364 i&13.96257983 + 8.45167276 i&14.54728640 + 9.28905550 i\\ \hline
7 &15.99974275 - 7.23543754 i&16.73291571 - 8.19691791 i&17.42277980 - 9.11081249 i\\ \hline
8 &15.99974275 + 7.23543754 i&16.73291571 + 8.19691791 i&17.42277980 + 9.11081249 i\\ \hline
9 &18.42897216 - 6.79355442 i&19.26164201 - 7.81977962 i&20.04405940 - 8.79717049 i\\ \hline
10&18.42897216 + 6.79355442 i&19.26164201 + 7.81977962 i&20.04405940 + 8.79717049 i\\ \hline
11&20.69785002 - 6.27781735 i&21.62130874 - 7.35928644 i&22.48793355 - 8.39101050 i\\ \hline
12&20.69785002 + 6.27781735 i&21.62130874 + 7.35928644 i&22.48793355 + 8.39101050 i\\ \hline
13&22.84567896 - 5.70854973 i&23.85362358 - 6.83821861 i&24.79844419 - 7.91745869 i\\ \hline
14&22.84567896 + 5.70854973 i&23.85362358 + 6.83821861 i&24.79844419 + 7.91745869 i\\ \hline
15&24.89759650 - 5.09866262 i&25.98519281 - 6.27108315 i&27.00361964 - 7.39255586 i\\ \hline
16&24.89759650 + 5.09866262 i&25.98519281 + 6.27108315 i&27.00361964 + 7.39255586 i\\ \hline
17&26.87086392 - 4.45689278 i&28.03424697 - 5.66771323 i&29.12262537 - 6.82719343 i\\ \hline
18&26.87086392 + 4.45689278 i&28.03424697 + 5.66771323 i&29.12262537 + 6.82719343 i\\ \hline
19&28.77796298 - 3.78943945 i&30.01394562 - 5.03509475 i&31.16927222 - 6.22911873 i\\ \hline
20&28.77796298 + 3.78943945 i&30.01394562 + 5.03509475 i&31.16927222 + 6.22911873 i\\ \hline \hline
\end{tabular}

\end{document}